\documentclass[10pt,letterpaper]{article}
\usepackage{opex3}
\usepackage{amsmath}
\usepackage{graphicx}
\usepackage{sistyle}
\usepackage{todonotes}
\usepackage{gensymb}

\usepackage{bm, dsfont}

\begin{document}

\clearpage

\title{Narrowband Photon Pair Source for Quantum Networks}
\author{F.~Monteiro, A.~Martin, B.~Sanguinetti*, H.~Zbinden and R.~T.~Thew}
\address{Group of Applied Physics, University of Geneva, Switzerland}
\email{Bruno.Sanguinetti@unige.ch}

\begin{abstract}
We demonstrate a compact photon pair source based on a periodically poled lithium niobate nonlinear crystal in a cavity. The cavity parameters are chosen such that the emitted photon pair modes can be matched in the region of telecom ultra dense wavelength division multiplexing (U-DWDM) channel spacings. This approach provides efficient, low-loss, mode selection that is compatible with standard telecommunication networks.  Photons with a coherence time of 8.6\,ns (116\,MHz) are produced and their purity is demonstrated. A source brightness of 134\,pairs\,(s.\,mW.\,MHz)$^{-1}$ is reported. The high level of purity and compatibility with standard telecom networks is of great importance for complex quantum communication networks. 
\end{abstract}

\ocis{(270.5565) Quantum communications; (190.4410) Nonlinear optics, parametric processes}

\bibliographystyle{osajnl}

\bibliography{cavity_bibliography4}

\section{Introduction}

Narrowband photon pair sources are essential for quantum communication networks~\cite{Gisin07} to provide robustness to fibre length fluctuations as well as polarisation and chromatic mode dispersion~\cite{Fasel04}. This is of critical importance for quantum communication primitives such as entanglement swapping, where two  photons that shared no previous correlations become entangled~\cite{Zukowski93}. Narrowband photons help to ensure a good overlap of the photons for the Bell state measurement~\cite{Halder2007}. A common way to generate narrowband photonic states for entanglement swapping experiments is by spectrally filtering the Spontaneous Parametric Down Conversion (SPDC) emission \cite{Tanzilli2010,Tanzilli2012}. To achieve high fidelity operation, we need to ensure that the photons are both indistinguishable and pure~\cite{Grice01,Osorio13}.

In the case of pulsed systems, the spectral filter bandwidth is chosen according to the coherence time of the pump pulse~\cite{Tanzilli2010}, and is around \SI{10}{GHz} for picosecond pulses. However, femtosecond pulsed systems are typically too broadband for practical applications in telecom networks. In the case of combining multiple sources based on SPDC pumped by continuous wave (CW) lasers, the purity depends on the detection jitter \cite{Halder2007,Kumar2010}, i.e. the detector jitter has to be smaller than the filtered photon's coherence time. Typical detector jitter when working at telecom wavelengths can vary between \SI{60}{} to \SI{400}{ps} \cite{Polyakov2011}, which requires filtering between \SI{0.1}{} to \SI{1}{GHz}. The challenge is thus to implement a low-loss and narrowband filtering solution. One should also note that even for a filter with a peak transmission of 1, a Gaussian filter will still have a loss of $ 1/\sqrt{2} $ for a perfectly Gaussian bandwidth-matched photon.
  
A promising approach to both narrowband photon generation and filtering is based on placing the nonlinear crystal in an optical cavity in an Optical Parametric Oscillator (OPO) configuration.~\cite{Benson2009,Pomarico2009,Moreno2010,Benson2011,Pomarico2012,Harris2012,Riedmatten2013,Silberhorn2013}.  OPO sources operating below threshold produce photons in the allowed cavity modes and their bandwidths can be tuned by changing the length and cavity mirror reflectivities~\cite{Moreno2010,Eckardt1991,Pomarico2012}. In addition, the  brightness of this kind of source is enhanced compared to the crystal without a cavity \cite{Ou2000,Sanders2000,Moreno2010}. However, selecting the desired photon pairs usually requires additional filtering to obtain only the desired cavity modes - OPO sources typically produce a frequency comb of many frequencies spaced according to the free spectral range (FSR) of the cavity. 
 
Due to the narrowband nature of the photons, the loss associated with selecting these modes will only depend on the peak transmission loss - most filters will be effectively "top hat" filters. However, due to the relatively small FSR of most of these sources~\cite{Riedmatten2013,Pomarico2009,Harris2012,Benson2009,Benson2011,Silberhorn2013}, another cavity normally has to be utilised for this mode selection. In this paper we present a compact OPO photon pair source that is designed to have a large FSR that is close to the ITU channel spacing of commercially available ultra dense wavelength division multiplexing (U-DWDM) technology. Coupled with the Telecom wavelength emission of our source, this provides a versatile low-loss narrowband source of photon pairs for multi-source, multi-photon, quantum communication networks and protocols.

\begin{figure}[htp]
\centering
\includegraphics[scale=0.48]{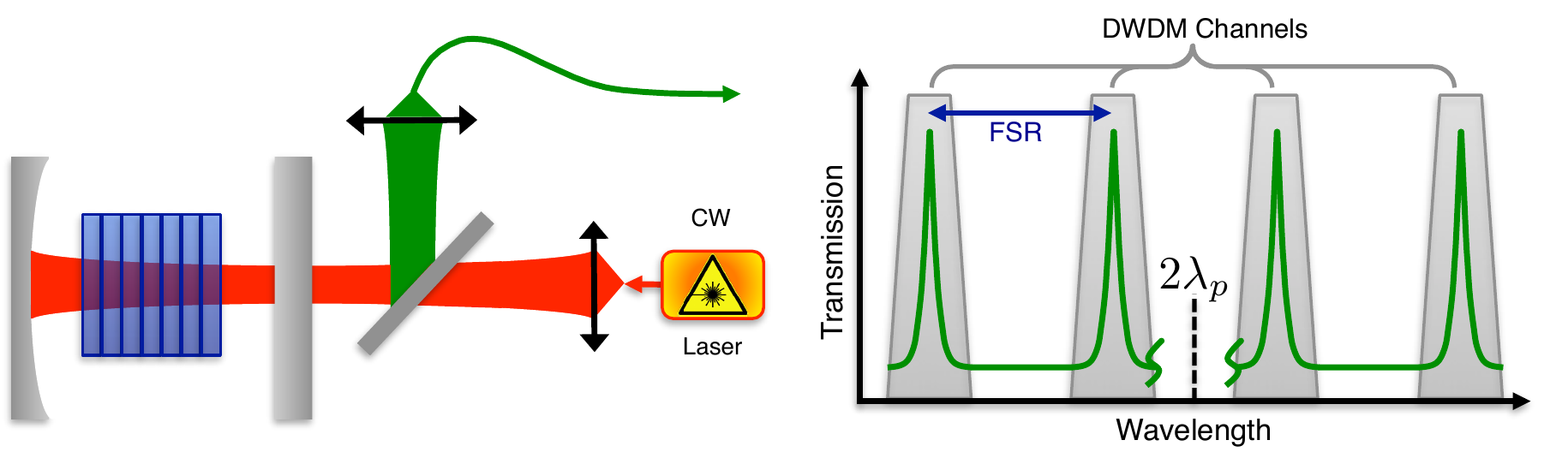}
\caption{(Left) OPO Photon pair source geometry. (Right) The short ($\sim$\,3.9\,mm) cavity length is chosen so that the FSR  matches the DWDM spacing to allow for low-loss mode selection.}
\label{concept}
\end{figure}

An OPO source is obtained by placing a nonlinear medium such as a bulk crystal~\cite{Harris2012,Benson2011,Moreno2010,Riedmatten2013} or a waveguide~\cite{Pomarico2009,Pomarico2012,Silberhorn2013} inside an optical cavity. Reflection coatings can be placed on the end-faces making for an efficient and compact device. In this case, we have chosen to use a geometry as depicted on the left of Fig~\ref{concept} to allow us to easily tune the cavity parameters while testing for an optimal configuration, allowing this set-up to be a proof of principle source that can be used in the design of future experiments that require narrow bandwidth, pure photon pair source that use several output channels of a U-DWDM.  

\section{Description and Set-up}

One of the constraints is to have a FSR of around \SI{25}{GHz}, close to an ITU U-DWDM channel spacing, as depicted on the right of Fig~\ref{concept}. This limits the size of our nonlinear crystal, a type zero MgO doped period poled lithium niobate crystal (Covesion), to 1.0\,mm in a 3.9\,mm cavity. In order to have a resonant down-converted beam, we use one flat and one concave mirror with a \SI{12}{mm} focal length, resulting in a waist of the resonant gaussian mode of \SI{65}{\mu m} at the middle of the cavity. Taking into account the mirror's curvature and the refractive index of the crystal this gives a FSR of \SI{29.7}{\pm}{1.0}{GHz}(\SI{0.241}{\pm}{0.008}{nm}).

\begin{figure}[here!]
\centering
\includegraphics[scale=0.14]{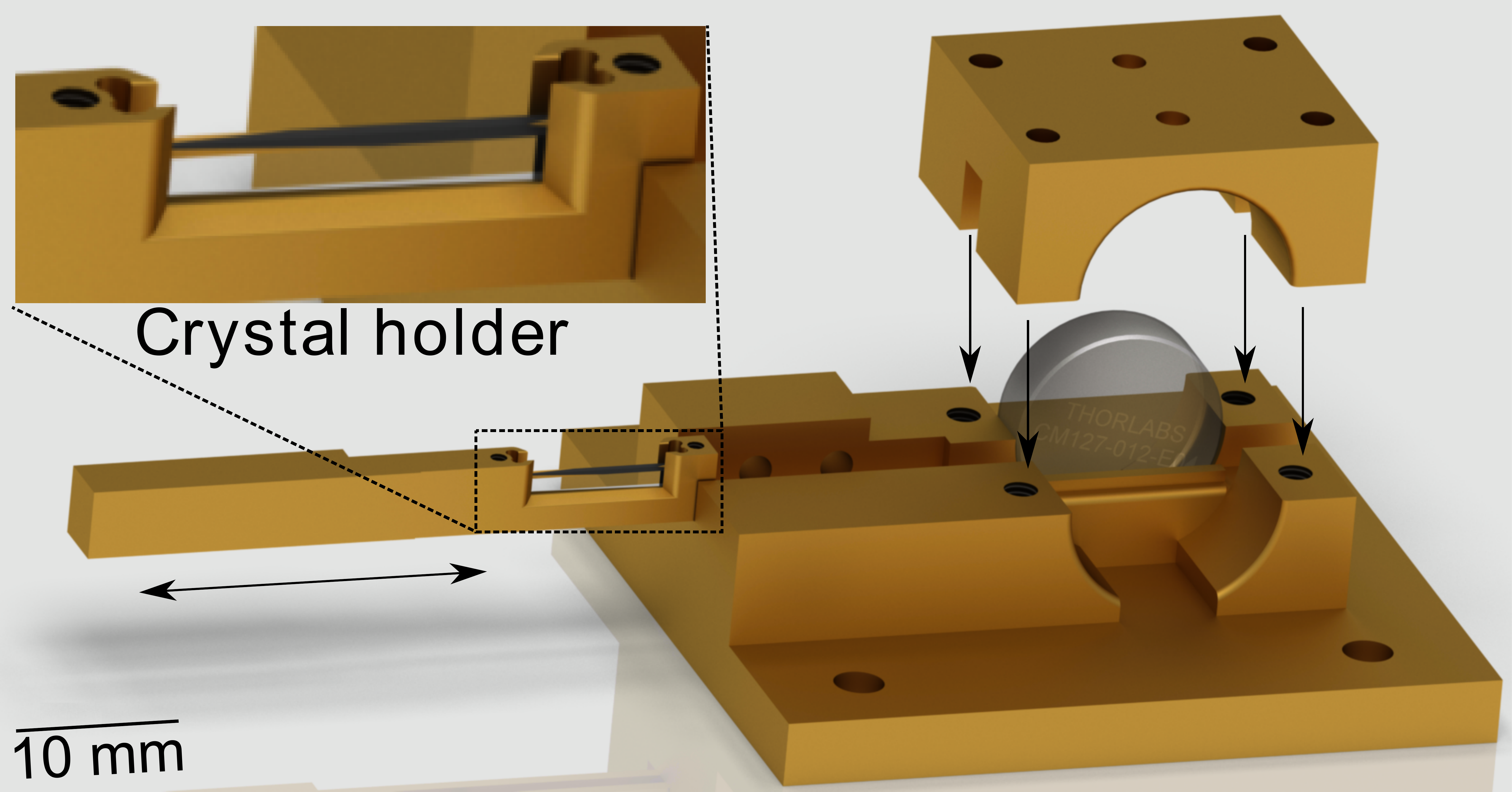}
\caption{OPO photon pair source cavity mount. The upper inset is a zoom on the crystal and holder.  The design is made in such a way that the mean height of the crystal coincides with the center of the concave mirror to within \SI{100}{\mu m}.}
\label{mount}
\end{figure}

The test cavity is shown in Fig~\ref{mount} and was designed to be sufficiently flexible so as to allow for the mirrors and crystal to be quickly changed without losing alignment. The cavity has mirrors that are reflective for telecom wavelengths and transparent for the pump wavelength. The crystal holder is designed to smoothly follow a track, allowing us to scan across several different poling periods in the crystal without losing the cavity alignment. The structure is made in such a way that the center of the crystal always coincides with the center line of the mirrors within a distance of less than \SI{100}{\mu m}. The temperature of the mount and crystal is controlled by a resistor fixed to the lower plate.

\begin{figure}[here!]
\centering
\includegraphics[scale=0.7]{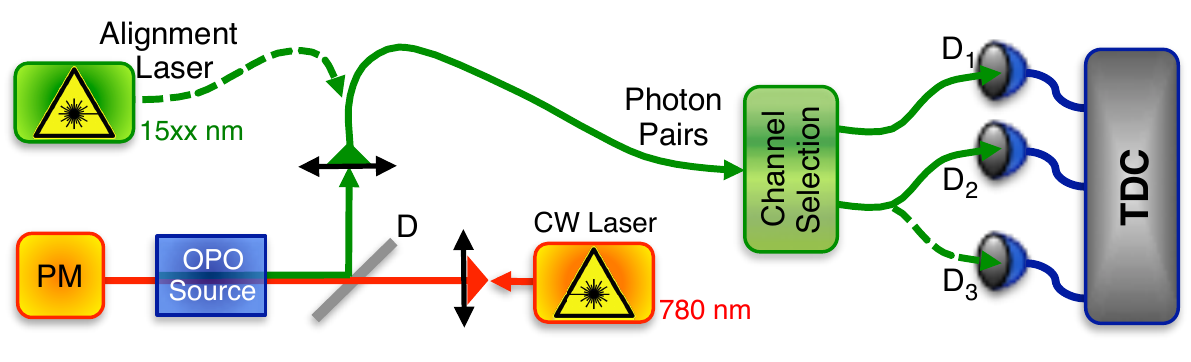}
\caption{Schematic of set-up for source characterisation. A power meter (PM) measures the transmitted pump power to characterise the cavity. A dichroic mirror (D) separates the emitted photon pairs from any pump light. The pairs are coupled into single mode Telecom fibres before deterministically separating the photons (channel selection). Three detectors are used to check the source characteristics and statistics with the aid of a Time-to-Digital Converter (TDC).}
\label{setup}
\end{figure}

The experimental set-up used to characterise the source is illustrated in Fig~\ref{setup}. A tunable, telecom-band, alignment laser (Tunics Nettest) and power meter (PM) are used to align and characterise the cavity. To generate the photon pairs, the source is pumped by a fiber-coupled CW laser (Toptica DL100), locked on the Rubidium transition at \SI{780.24}{nm}.  A dichroic mirror (D) separates the emitted photon pairs from any pump light in the reflected mode. The pairs are coupled into single mode telecom fibres before deterministically separating the photons (channel selection). In this first instance we use one tunable gaussian filters (JDS TB9) with \SI{24.6}{GHz} pass band and one DWDM, before detection and analysis. Various combinations of pairs of single photon detectors (SPDs) are used for different measurements. In the following we elaborate on the alignment and characterisation measurements.

\section{Measurements and Characterization}

\subsection{Free Spectral range}

\begin{figure}[htp]
\centering
\includegraphics[scale=0.4]{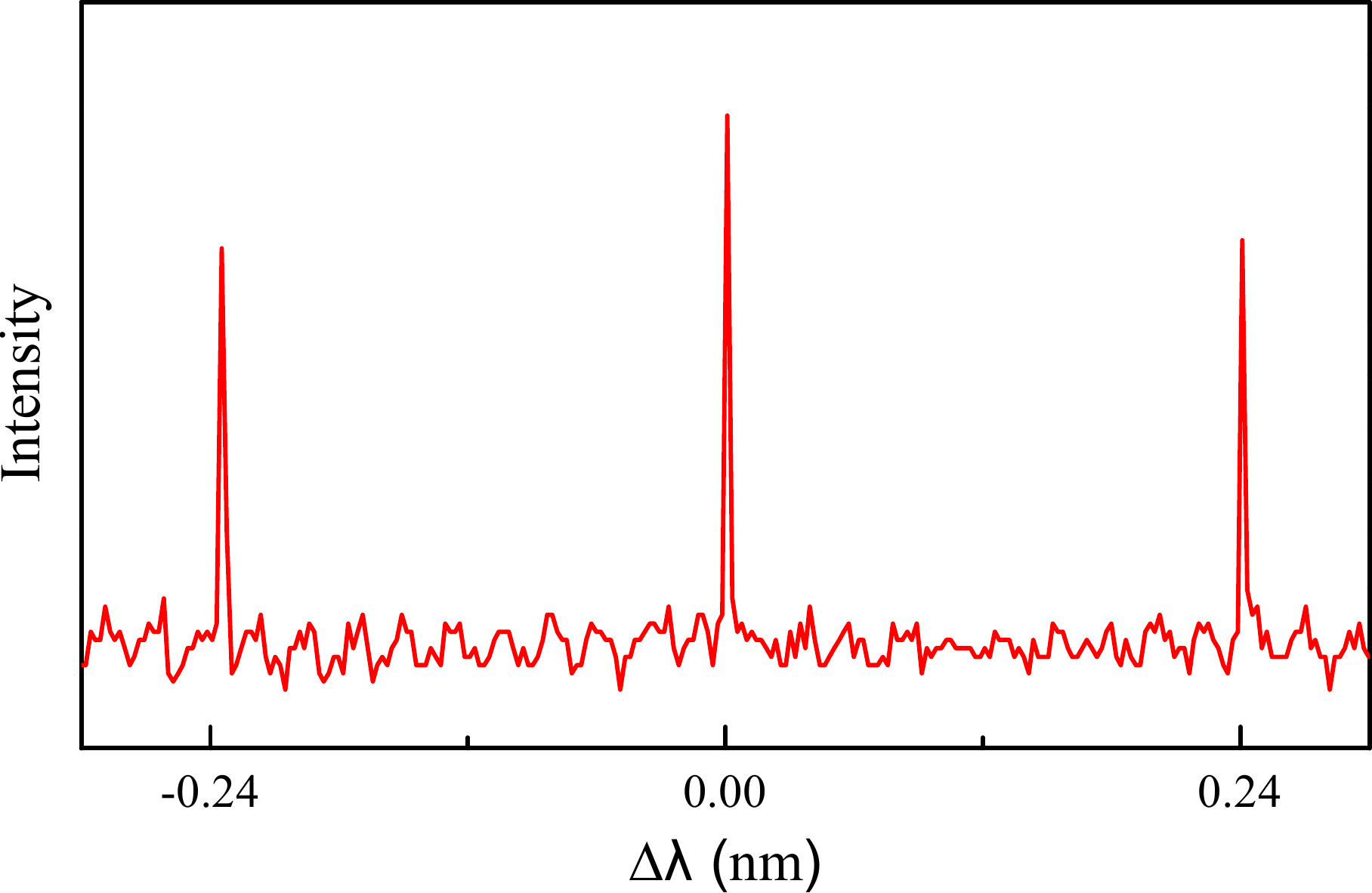}
\caption{Measured cavity FSR. We obtain a FSR of \SI{0.24}{nm} (\SI{29.7}{GHz}), which is in agreement with the expected value. The lack of other cavity modes shows a good alignment between the cavity and the single mode telecom fiber.}
\label{FSR}
\end{figure}

The FSR is measured by injecting the tunable telecom-band laser beam backwards through the system. In Fig~\ref{setup}, the alignment laser is sent into the cavity via the dichroic mirror (D). A \SI{15}{cm} focal length lens is chosen to match the \SI{65}{\mu m} waist of the cavity. A power meter (PM) measures the transmission through the cavity as the laser wavelength is scanned around 1560\,nm. Fig~\ref{FSR} shows that the measured FSR around \SI{1560}{nm} is \SI{0.24}{nm}, which is in agreement with the expected value.

\subsection{Generated Photon pair spectrum}

The photon pair spectrum was measured using a spectrometer composed of a grating and an InGaAs CCD sensor sensitive at the single photon level. In Fig 5 (Left) we see the spectrum for the emitted photon pairs at a specific cavity temperature (\SI{43.77}{^{\circ}C}). Due to the spectrometer resolution, which is close to \SI{0.5}{nm}, it is not possible to see the detailed spectral structure of the cavity itself, but only its envelope. There is also a slight asymmetry due to the lower detection efficiency for high wavelengths in this range.

\begin{figure}[htp]
\centering
\includegraphics[scale=0.31]{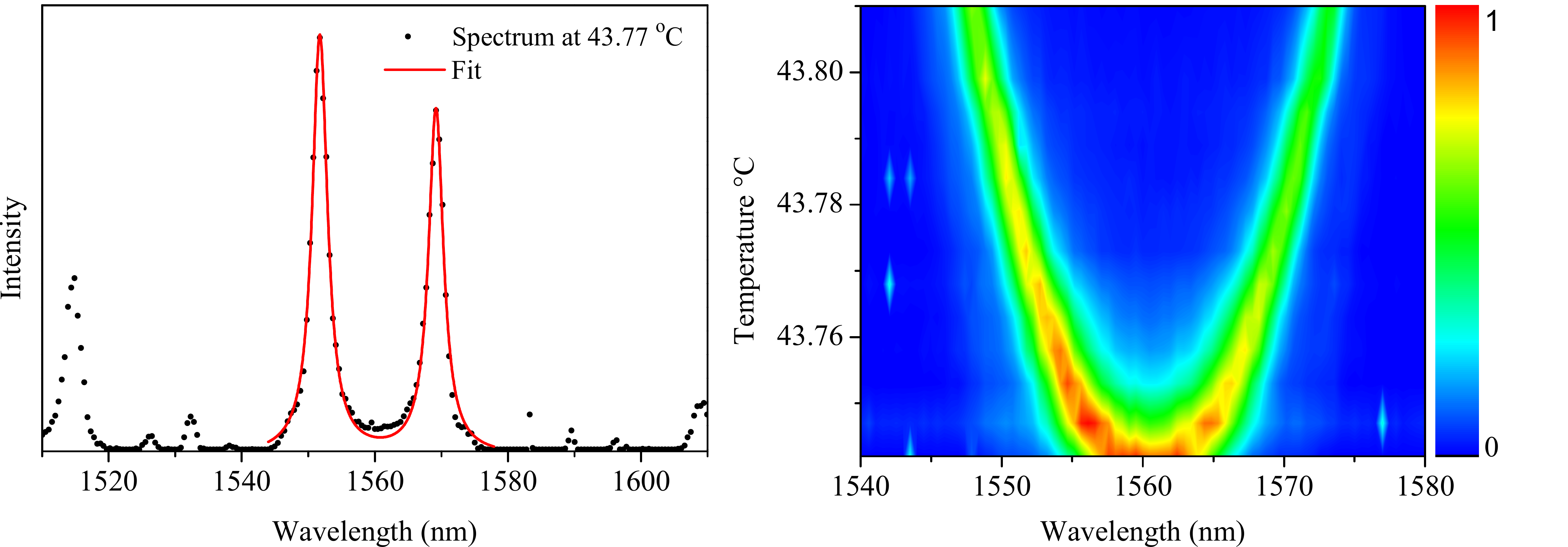}
\caption{Left: The measured spectrum with the cavity at 43.77$ ^{o} $C is dominated by two clusters centered around $ 2\lambda_{P} $. At this temperature, the width of the cluster envelope is around \SI{2.5}{nm}. The decrease in height with the increasing wavelength is due to a decrease in the sensitivity of the InGaAs sensor with respect to the wavelength. Right: Envelope position of the clusters with respect to the temperature of the crystal and cavity.}
\label{spectrum}
\end{figure}

Typically, for such a short crystal without cavity, one would expect to see this spectral envelope extend across the entire wavelength range shown in fig~\ref{spectrum}. What we see here is an effect of chromatic dispersion for the signal and idler photons. The different refractive indicies for the different wavelengths $ n_{s,i} $ will result in what is called the clustering effect \cite{Eckardt1991,Pomarico2009,Pomarico2012}, in which the OPO's spectrum is composed of photon pairs that not only satisfy the energy conservation $ \omega_{s} $ , $ \omega_{i} = \omega_{pump} - \omega_{s} $ and phase-matching conditions, but also satisfy the allowed cavity modes.

The ability to tune the wavelength position of the cluster, and the fact that we have several cavity modes inside each envelope peak, allows us to fine-tune the spectral alignment of the cavity modes to a DWDM channel or filter for mode selection. The  peak position of these clusters can be tuned by changing the effective optical path-length, e.g. by changing of the temperature of the crystal and cavity \cite{Eckardt1991}. Fig~\ref{spectrum} (Right) shows the behaviour of the position of the envelope peaks of Fig~\ref{spectrum} (Left) as a function of the crystal and cavity temperature, which shows that we can tune across the entire telecom C-band and hence all DWDM channels.

Once the source is emitting photons that are spectrally aligned to the filters, we use the SPDs and the time to digital converter (TDC), to characterise the correlated photon pairs. The filter bandwidth is much larger than that of the expected photons, but smaller than the FSR.  This allows us to select the desired photon pair modes with low loss. As such when the TDC measures the arrival time difference between the two photons the width of the resulting coincidence peak directly provides us with information on the bandwidth of the emitted photons.  In Fig~\ref{setup},  we saw that the pump beam enters the cavity through a flat mirror and the downconverted photons escape back through the same mirror. We measured a reflectivity of $ \sim $ 0.995 and 0.985 at 1560 nm for the concave and flat mirrors respectively. The measured transmission of the crystal is $ \sim 0.997 $, leading to an expected finesse of 243 and a expected photon coherence time of \SI{\sim 8.2}{ns}.

\subsection{Coincidence Measurement}

In practice, we first select one cavity mode with a standard DWDM and then maximise the singles count rates on a free running detector (D$_1$) (IDQ - id220; 15\% efficiency, \SI{600}{Hz} of noise, \SI{20}{\mu s} deadtime) placed after a tunable filter with 24.6 GHz bandwidth, which mimics the U-DWDM. The coincidence measurements are then made between (D$_1$), and a second triggered SPD (D$_2$) (IDQ - id210; 20\% efficiency and \SI{20}{ns} gate) which is aligned with the other DWDM channel. 

\begin{figure}[htp]
\centering
\includegraphics[scale=0.31]{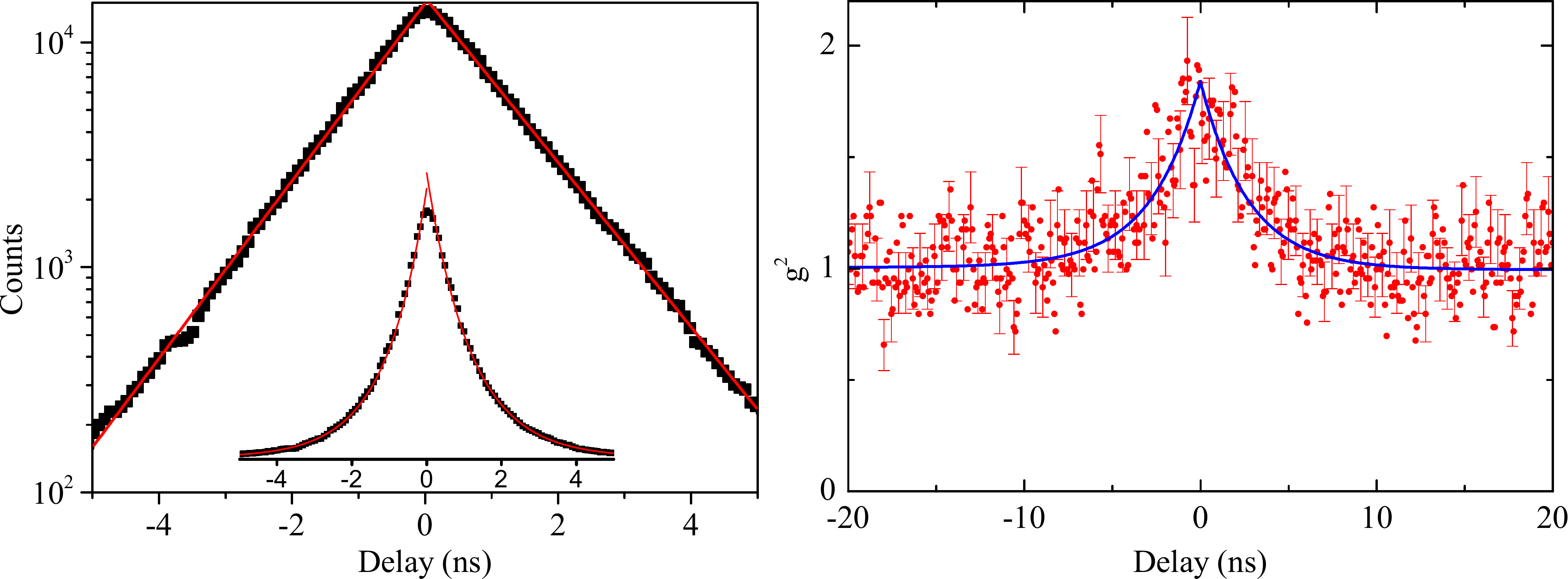}
\caption{Left: Coincidence measurement for the two correlated cavity modes. The central inset shows this measurement in the linear scale. The noise to signal ratio is less than 1\%. Right: The g$ ^{(2)} $ measurement of our source. We show the error bars at every four points in order to clarify the plot. The measured effective number of modes \cite{Silberhorn2012} is 1.19 $ \pm $ 0.15.}
\label{coin}
\end{figure}

Fig~\ref{coin} (Left) shows the coincidence peak on a logarithmic scale, and with a linear fit, in order to highlight the exponential decay typical of this kind of source \cite{Harris2012,Benson2011,Moreno2010,Silberhorn2013,Silberhorn2012,Riedmatten2013}. By fitting the decay with $ e^{\pm2\pi\delta fT} $ \cite{Riedmatten2013}, where $ T $ is the delay, we find a bandwidth of $ \delta f = $ \SI{116}{MHz} for both photons, which corresponds to a Finesse of 255, in good agreement with the expected value. Given the lorentzian shape of the cavity mode, this bandwidth gives a coherence time of 8.6 ns, which is much larger than our detector jitter (\SI{\sim 300}{ps}). An error of less than 1\% for the measurement of the mirror reflectivities and and crystal  characteristics can account  for the discrepancy between the measured and expected finesse values. Correcting for detection efficiencies and the external losses, the measured brightness of the source is 134\,pairs\,(s.\,mW.\,MHz)$^{-1}$, which is enhanced by a factor of $ \sim $ 500 when compared to the crystal without the cavity. We obtain a photon pair creation probability, with  40\,mW of pump power, of around $5\,\times\,10^{-3}$ pairs per mode. The adaptation of these OPO characteristics to a short PPLN waveguide device~\cite{Pomarico2012} would allow much higher values for even lower pump powers.

By analysing the coincidences and singles rates, we find a source-fiber coupling efficiency of 25\%. Given the mirror reflectivities and crystal transmission described above, we estimate that the escape probability of a photon through the flat mirror is $ \sim $ 45\%, which implies that the free space to fiber coupling efficiency is $ \sim $ 60\%.

\subsection{Source Purity}

In multi-photon experiments such as entanglement swapping, one of the most important characteristics for the photons is their purity~\cite{Grice01,Osorio13}. In the pulsed regime this is much clearer than for CW pumped systems. In the case of CW systems, the role of detection is analogous to that of the pump laser bandwidth in the pulsed regime. This was already highlighted in a previous experiment by some of us~\cite{Halder07} where the visibility of a HOM interference dip between independent sources was dependent on the detector jitter being smaller than the photon coherence time. This was recently put on a more rigorous footing~\cite{Kumar2010}. In our case, we have a coherence time of \SI{8.6}{ns} and the detectors used to measure the purity have a jitter of \SI{0.2}{ns}.  

To characterise the purity, spectrometers are typically used to measure the joint spectral intensity. Given the narrowband nature of our photons this is beyond the resolution of our spectrometers. An alternative approach, however, has shown that this information can be extracted from a measurement of g$ ^{(2)} $(0)~\cite{Silberhorn2012,Bruno13}. Specifically, we can use the relationship g$ ^{(2)} (0) = 1 - 1/N $, where $ N $ is the effective number of modes. To measure this, we use a 50/50 beam splitter  placed after the mode selection, see Fig~\ref{setup}, and the coincidences were recorded using two custom low-noise free running SPDs \cite{Boris2013}, D$_2$ and D$_3$. Fig~\ref{coin} (Right) shows the g$ ^{(2)} $ measurement. We find a g$ ^{(2)}(0) $ = 1.84 $ \pm $ 0.11, which gives $ N $ = 1.19 $ \pm $ 0.15, indicating a high level of purity for our system.

\section{Conclusion}

We have presented a simple and compact photon pair source that is both well suited to producing narrowband photons and interfacing with telecom DWDM technologies. The high level of purity and compatibility with standard Telecom networks is of great importance for complex quantum communication networks. In the future we plan to apply the methods described here to the design of a monolitical device with \SI{25}{GHz} of FSR, consisting of a waveguide coated at the two sides. This will be further applied in future experiments that require low filtering losses, photons with long coherence time, high level of purity and photons at several output channel of a U-DWDM.

\section{Acknowledgements}

The authors would like to thank Boris Korzh for assistance with the low noise free running  detectors and we acknowledge the Swiss NCCR QSIT for financial support.

\end{document}